\documentclass[12pt,a4paper]{article}
\usepackage{latexsym}
\usepackage{psfig}
\usepackage{rawfonts}
\newcommand{\nsz}{\normalsize}

\newcommand{\beq}[1]{\large\begin{equation}\label{#1}}
\newcommand{\eeq}{\end{equation}\nsz}
\newcommand{\bma}{\large\[}
\newcommand{\ema}{\]\nsz}

\newcommand{\Eq}[1]{(\ref{#1})}
\def\bear{\begin{eqnarray}}
\def\ear{\end{eqnarray}}

\begin{document}

\vspace*{1.0cm}

\centerline{\LARGE Antiprotons Annihilation in the Galaxy}
\centerline{{\LARGE As A Source of Diffuse Gamma Background}
\footnote{Submitted to Yadernaya Fizika}}

\bigskip
\centerline{\large Yu.A.~Golubkov{\small $^{a)}$},
M.Yu.~Khlopov{\small $^{b)}$}}

\bigskip
\centerline{D.V.Skobeltsyn Institute of Nuclear Physics}
\centerline{M.V.Lomonosov Moscow State University}

\bigskip
\centerline{\it {\small $^{a)}$}Institute of Nuclear Physics,
Moscow State University, Vorobjevy Gory,}
\centerline{\it 119899, Moscow, Russia}
\centerline{\it {\small $^{b)}$}Center for CosmoParticle Physics COSMION,}
\centerline{\it Miusskaya Pl.4, 125047, Moscow, Russia}


\begin{abstract}%
The existence of antimatter domains in baryon asymmetrical Universe
can appear as the cosmological consequence of particle theory
in inflationary models with non-homogeneous baryosynthesis.
Such a domain can survive in the early Universe and form
globular cluster of antimatter stars in our Galaxy.
The model of antimatter pollution of Galaxy and annihilation with 
matter gas is developed.
The proton-antiproton annihilation gamma flux is shown to reproduce
the observed galactic gamma background measured by EGRET. 
From comparison with observational data the estimation on the maximally
allowed amount of antimatter stars, possibly present in our Galaxy,
is found.
\end{abstract}



\newpage
\section{Introduction}

The generally accepted motivation for baryon asymmetric 
Universe is the observed absence of the macroscopic amounts of antimatter 
up to the scales of clusters of galaxies, which probably extends 
on all the part of the Universe within the modern cosmological horizon 
\cite{Cohen98}.
The modern cosmology relates this baryon asymmetry of the Universe
to the process of baryosynthesis., i.e. to the creation of baryon excess
in very early Universe \cite{Sakharov67,Kuzmin70}. 
In the homogeneous baryon asymmetric Universe the Big Bang theory predicts 
exponentially small fraction of primordial antimatter.
Therefore, any non exponentially small amount of antimatter 
in the modern Universe is the profound signature for new phenomena, 
related to the existence of antimatter domains
and leads to the respective predictions for antinuclear 
component of galactic cosmic rays.

The most recent analysis finds that the
size of possible antimatter domains in baryon symmetrical Universe
should be only few times smaller 
than the modern cosmological horizon to escape the contradictions 
with the observed gamma ray background \cite{Cohen98}. 
The distribution of antibaryon excess, 
corresponding to relatively small ($<\,10^{-5}$) volume occupied by it, 
can arise in inflational models with baryosynthesis and is compatible 
with all the observational constraints on the annihilation of antimatter
in the baryon dominated Universe \cite{Khlopov00}. 
The size and amount of antimatter domains
is related to the parameters of models of inhomogeneous
baryosynthesis (see for review \cite{Chechetkin82,Khlopov96}).
With the account for all possible mechanisms for 
inhomogeneous baryosynthesis, predicted on the base of
various and generally independent extensions of the standard
model, the general analysis of possible domain distributions
is rather complicated. But the main point of the existing mechanisms 
of baryosynthesis, important for our aims, is that all of them
can lead to inhomogeneity of baryon excess generation
and even to generation of antibaryon excess in some regions of space,
when the baryon excess, averaged over the whole space, being positive
(see reviews in \cite{Chechetkin82,Khlopov99,Chechetkin87}).  

On the other hand, EGRET data \cite{EGRET97} on diffuse gamma 
background show visible peak around $E_{\gamma}\,\approx\,70$ MeV
in gamma spectrum, which fact can be naturally explained by
the decays of $\pi^0$-mesons, produced in nuclear reactions.
Interactions of the protons with gaseous matter in the Galaxy shift
the position of such a peak to higher values of gamma energy
due to $4$-momentum conservation.
At the same time the secondary antiprotons, produced 
in the cosmic ray interactions with interstellar gas, 
are too energetic \cite{Chardonnet96} and their annihilation
also cannot explain the observational data.

The above consideration draws attention to the model with 
antimatter globular cluster existing in our Galaxy, 
which cluster can serve as a permanent source of antimatter
due to (anti)stellar wind or (anti)Supernova explosions.
The isolated antimatter domain can not form astronomical object 
smaller than globular cluster \cite{Khlopov98}. 
The isolated anti-star can not be 
formed in the surrounding matter since its formation implies the 
development of thermal instability, during which cold clouds are 
pressed by hot gas. Pressure of the hot matter gas on the antimatter 
cloud is accompanied by the annihilation of antimatter. 
Thus anti-stars can be formed in the surrounding antimatter only, 
what may take place when such surrounding has at least 
the scale of globular cluster.
One can expect to find antimatter objects among 
the oldest population of the Galaxy \cite{Khlopov98},
in the halo, since owing to strong annihilation of 
antimatter and matter gas the formation of secondary antimatter objects 
in the disk component of our Galaxy is impossible. 
So in the estimation of antimatter effects we can use the data 
on the spherical component of our Galaxy as well as 
the analogy with the properties of the old population stars 
in globular clusters and elliptical galaxies.
The total mass of such cluster(s) is constrained from below 
by the condition of antimatter domain survival in the surrounding 
baryonic matter because small antimatter domains completely annihilate 
in the early Universe before the stage of galaxy formation.
The upper limit on the total mass of antimatter can be estimated
from the condition, that the gamma radiation 
from annihilation of antimatter
with galactic matter gas does not exceed the observed galactic 
gamma background.
The expected upper limit on cosmic antihelium flux from antimatter stars 
in our Galaxy was found \cite{Khlopov98,Golubkov00}
only factor of two below the modern level 
of sensitivity in experimental cosmic antihelium searches \cite{AMS}.
In the first approximation the integral effect we study depends 
on the total mass of the antimatter stars and does not depend 
on the amount of globular clusters. The only constraint is 
that this amount does not exceed the observed number
of galactic globular clusters (about 200).

Assume that antimatter globular
cluster, moves along elliptical orbit in the halo. 
The observed dispersion of velocity of globular clusters is
$<v>\,\sim\,300$ km/s and of the long axis of their orbits is
$<r>\,\sim\,20$ kpc. This gives $T\,\sim\,2\cdot 10^{15}$ s as the order 
of the magnitude for the period of orbital motion of the cluster
in the Galaxy. The period the cluster moves along the dense region 
of the disk with the mean half--width $D\,\sim\,100$ pc depends 
on the angle at which the orbit crosses the plane of the disk 
and is of the order

\bma
t_d\ \sim\ \frac{D}{<v>}\ \sim\ 10^{13}\ s.
\ema

This means that the cluster spends not more than 1\% 
of the time in the dense region of galactic disk, where 
the density of gas is of the order of $n_H^{disk}\,\sim\,1$ cm$^{-3}$, 
moving the most time in the halo with much lower density 
of the matter gas $n_H^{halo}\,\sim\,5\cdot 10^{-4}$ cm$^{-3}$.
Therefore, we can neglect the probability to find the cluster
in the disk region and consider the case when the source of the antimatter
is in the halo.

One could expect two sources of the annihilation gamma emission from
the antimatter globular cluster.
The first one is the annihilation of the matter gas captured 
by the antimatter stars. Another source
is the annihilation of the antimatter, lost by the antimatter stars,
with interstellar matter gas. 
It is clear that the gamma flux originating from the annihilation 
of the matter gas on the antimatter stars surface is negligible.
Really, an antimatter star of the Solar radius $R\,=\,R_{\odot}$ 
and the Solar mass $M\,=\,M_{\odot}$ captures matter gas 
with the cross section

\bma
\sigma\ \sim\ \pi\,R\,\left (R\,+\,\frac{2GM}{v^2}\right )
\ \sim\ 4\cdot10^{22}\ cm^2\,,
\ema

\noindent so that the gamma luminosity of cluster of $10^5$ 
stars does not exceed $L_{\gamma}\,\le\,M_{5}\cdot 10^{29}\,erg/s$, 
where $M_5$ is the relative mass of the cluster in units $10^5\,M_{\odot}$,
$M_{cl}\,=\,M_5\cdot 10^5\,M_{\odot}$.
Such a low gamma luminosity being in the halo at the distance of about
$10$ kpc results in the flux $F_{\gamma}\,\le\,10^{-13}$ (ster$\cdot$
cm$^2\cdot$ s)$^{-1}$ of $1000$ MeV gamma rays near the Earth, 
what is far below the observed background. 
This explains why the antimatter star itself 
can be rather faint gamma source elusive for gamma astronomy and
shows that the main contribution into galactic gamma radiation may
come only from the annihilation of the antimatter lost 
by the antistars with the galactic interstellar gas.

There are two sources of an antimatter pollution from the (anti-)cluster: 
the (anti-)stellar wind and the antimatter Supernova explosions. 
In both cases the antimatter is expected to be spread out over the Galaxy 
in the form of positrons and antinuclei.
The first source provides the stationary in-flow of 
antimatter particles with the velocities in the range from
few hundreds to few thousands km/s to the Galaxy. 
The (anti)Supernova explosions give antimatter flows with velocities
order of 10$^4$ km/s. The relative contributions of both these
sources will be estimated further on the base of comparison
with the observational data assuming that all the contribution
into diffuse gamma background comes from the antimatter annihilation
with the interstellar matter gas.
We assume in present paper that the chemical content to be dominated 
by anti-hydrogen and consider the contribution 
from the annihilation of the antiprotons only. 

We consider the quasi-stationary case, provided
by the presence of a permanent source of the antimatter. 
The assumption about stationarity
strongly depends on the distribution of magnetic 
fields in the Galaxy, trapping charged antiparticles, annihilation
cross section and on the distribution of the matter gas.
We shall see that the assumption about stationarity
is well justified by existing experimental data and theoretical
models.

We carried out a careful consideration of the possibility 
to reproduce the observed spectrum of diffuse gamma background,
suggesting the existence of maximal possible amount of the antimatter
in our Galaxy. 
We showed that the predicted gamma spectrum is consistent 
with the observations. In this case the integral amount 
of galactic antimatter can be estimated, which estimation leads 
to definite predictions for cosmic antinuclear fluxes
\cite{Khlopov98,Golubkov00}, 
accessible for cosmic ray experiments in the nearest future \cite{AMS}.

\section{The model of galactic antimatter annihilation.}
In this section we shall show that one can consider
the antiproton annihilation in the halo as a stationary process
and the distribution of the antiprotons does not depend practically
on position and motion of the globular cluster of antistars.

One of the most crucial points for the considered
model is the annihilation cross section of the antiprotons.
In difference to the inelastic cross section of the $pp$
collisions, the cross section in the $\bar pp$ annihilation
steeply grows as kinetic energy of the antiprotons goes
to zero. This growth leads to the obvious fact that 
the main contribution into gamma flux must come from
the annihilation of the slowest antiprotons.
Therefore  we need to have reliable estimation for
the annihilation cross section of the antiprotons at low kinetic energies.
Existing theoretical models based mainly on the partonic picture 
of the hadronic interactions are definitely invalid 
for $\bar pp$ annihilation at low energies and
we used experimental data both for evaluation 
of the annihilation cross section as well as 
for the final state configuration.

At small energies the cross section must be proportional
to the inverse power of the antiproton velocity.
To find this dependence we have to match the available
experimental data on $\sigma_{ann}$ with this expected behavior.
As it follows from data \cite{Bertin96,Bruckner90}, obtained
at CERN-LEAR,  the dependence $\sigma_{ann}\,\sim\,v^{-1}$ 
is valid already for laboratory 
antiproton momenta $p_{lab}\,\le\,1000$ MeV/c.
The annihilation cross section is the difference between total 
and inelastic ones, $\sigma_{ann}\,\approx\,\sigma_{tot}-\sigma_{el}$.
Thus, at $P_{lab}\,\ge\,300\ MeV/c$ we used data from \cite{PDG96} 
for the total and elastic cross sections and
at momenta less than $300$ MeV/c we used the dependence

\beq{sigann}
\begin{array}{lll}
\sigma_{ann}(P\,<\,300\ \mbox{\nsz MeV/c}) & = & \sigma_0\,C(v^*)/v^*\,\\
\sigma_{el} & = & const\,,
\end{array}
\eeq

\noindent for annihilation and elastic cross sections, 
respectively, where $v^*$ is the velocity
of the antiproton in the $\bar pp$ center-of-mass system.
Additional Coulomb factor $ C(v^*)$ gives large increase 
for the annihilation cross section at small velocities of the antiproton
and is defined by the expression \cite{Landau3}:

\beq{coulfac}
C(v^*)\ =\ \frac{2\,\pi\,v_c/v^*}
{1\,-\,\exp\left ( -2\,\pi\,v_c/v^*\right )}\,,
\eeq

\noindent where, $v_c\,=\,\alpha\,c$, with $\alpha$ and $c$ being 
the fine structure constant and the speed of light, respectively.

Using the experimental data on the $\bar pp$ annihilation 
cross section \cite{Bertin96,Bruckner90} we found that value 
$\sigma_0$ in \Eq{sigann} is equal to:

\bma
\sigma_0 \ = \ \sigma_{ann}^{exp}(P\,=\,300\ \mbox{\nsz MeV/c})
\ =\ 160\ \mbox{\nsz mb}\,.
\ema

We used the spherical model for halo with $z$ axis directed to North Pole
and $x$ axis directed to the Solar system.
We parametrized the number density distribution 
of interstellar hydrogen gas $n_H(r,z)$ along $z$ direction as:

\beq{dskdens}
\begin{array}{lll}
n_H(z) & = & n_H^{halo}\,+\,\Delta_H(z)\,,\\
\\
\Delta_H(z) & = & \frac{n_H^{disk}}{1\,+\,(z/D)^2}\,,
\end{array}
\eeq

\noindent with $n_H^{halo}\,=\,5\cdot 10^{-4}$ cm$^{-3}$ being the hydrogen
number density in the halo, $n_H^{disk}\,=\,1$ cm$^{-3}$ being the hydrogen
number density in the disk and $D\,=\,100$ pc being the half-width of the
gaseous disk. We chose here the hydrogen number density in the halo
in suggestion that $\sim\,90\%$ of the halo mass is a non-baryonic dark matter.
Such a distribution of the matter gas is to large extent the worst
case for our aims since the matter density along $z$ axis falls
slowly and visible fraction of the antiprotons will annihilate sufficiently
far of the galactic disk plane. Nevertheless, as we shall see, even
in this case the picture is still quasi-stationary and the antiproton
number density in the halo is practically not disturbed by
the annihilation in the dense regions.

\begin{figure}[htb]                
\par
\centerline{\hbox{%
\psfig{figure=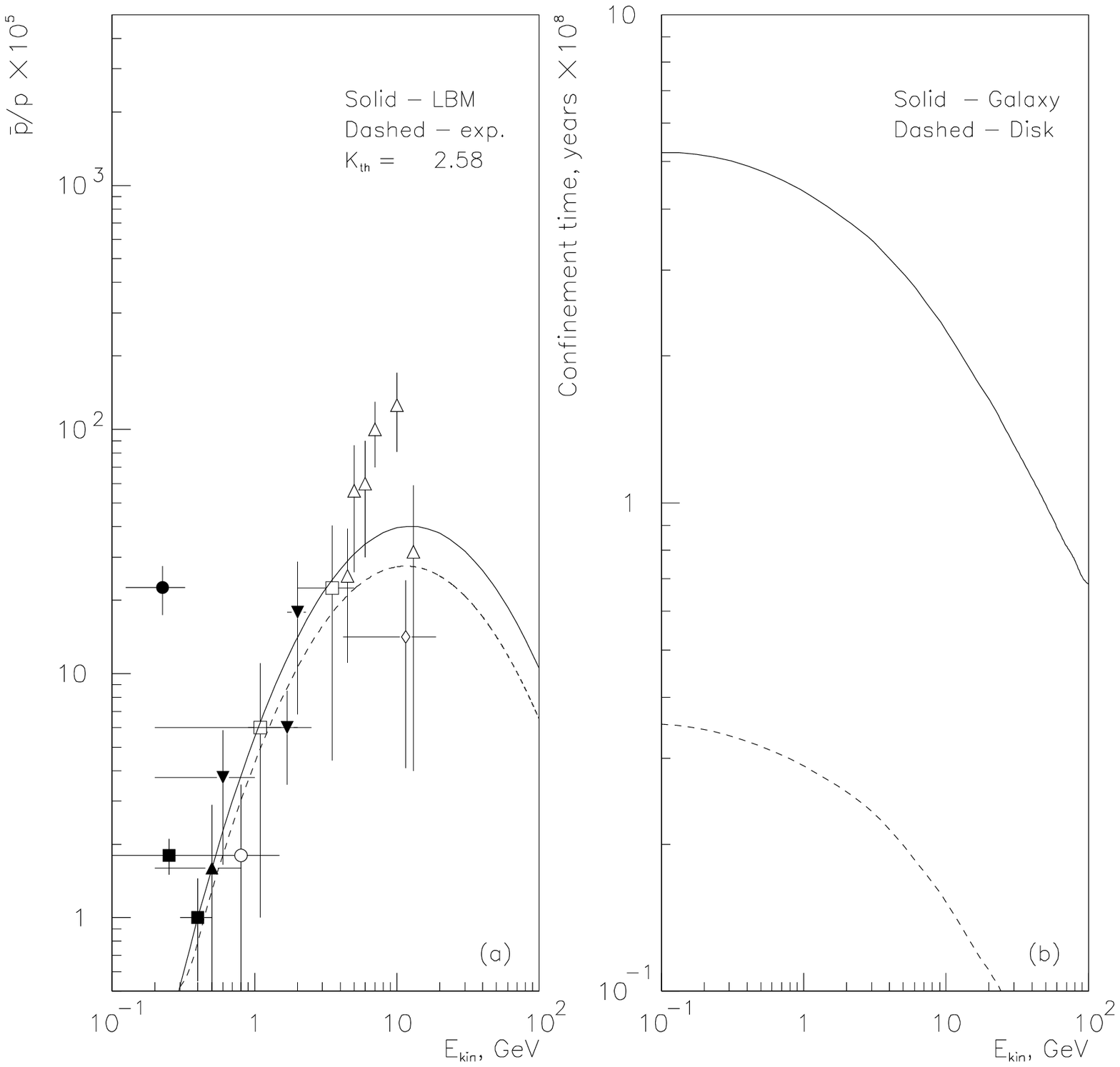,bbllx=1.5cm,bblly=5.5cm,%
bburx=19.0cm,bbury=23.0cm,clip=t,height=16.0cm}%
}}
\par
\caption{\label{expfit}
(a) Fit of the $\bar p/p$ ratio to experimental data.
Solid line shows predictions of the two--zone leaky box model
\cite{Chardonnet96}, increased by factor $K\,\approx\,2.6$. 
Dashed curve is the phenomenological fit, described in the text.
(b) The respective confinement times for the antiprotons 
in the Galaxy (solid) and in the disk (dashed).
The curves are taken from \cite{Chardonnet96} and multiplied by factor $K$.
}
\end{figure}

The validity of the stationary approximation depends on the interplay
of the life-time of the antiprotons to the annihilation
and their confinement time in the Galaxy.
To evaluate the antiproton confinement time we used the results 
of the ''two--zone'' leaky box model (LBM) \cite{Chardonnet96}.
The authors of \cite{Chardonnet96} considered the spectra of secondary
antiprotons produced in collisions of the cosmic ray protons with
interstellar gas. If to compare the antiproton spectrum, obtained
in \cite{Chardonnet96}, one easily observes that shape of the spectrum
beautifully reproduces the observational data on $\bar p/p$ ratio.
But the predicted total normalization is lower
by factor $2\div 3$ than the data.
Owing to the fact that confinement time enters as a common factor 
in the predicted $\bar p/p$ ratio, we found necessary factor, 
performing the fit to the observational data.
Experimental points have been taken from \cite{Golub98} 
where references on the data can be found.
The data on $\bar p/p$ ratio we used have been collected 
in balloon experiments and region of low kinetic energies, 
$E_{kin}\,\le\,100$ MeV, is strongly affected by the heliosphere 
\cite{Geer98}. To avoid this influence we removed from the fit 
two the most left points in Fig.\ref{expfit}. 
Solid curve in Fig.\ref{expfit}(a) represents
the ''two-zone'' LBM predictions for the $\bar p/p$ ratio,
multiplied by the fitted factor $K\ =\ 2.58$, which factor increases 
the confinement time for slow antiprotons in the Galaxy up to 
$5.5\cdot 10^8$ years. Dashed curve is the phenomenological fit in the form 
$R(E)\,=\,a\,E^{b+c\lg E}$, which we plotted for comparison.
The shapes of both curves match fairly.
Fig.\ref{expfit}(b) shows the resulting antiproton confinement times
for Galaxy as whole (solid) and for disk only (dashed).

\begin{figure}[htb]                
\par
\centerline{\hbox{%
\psfig{figure=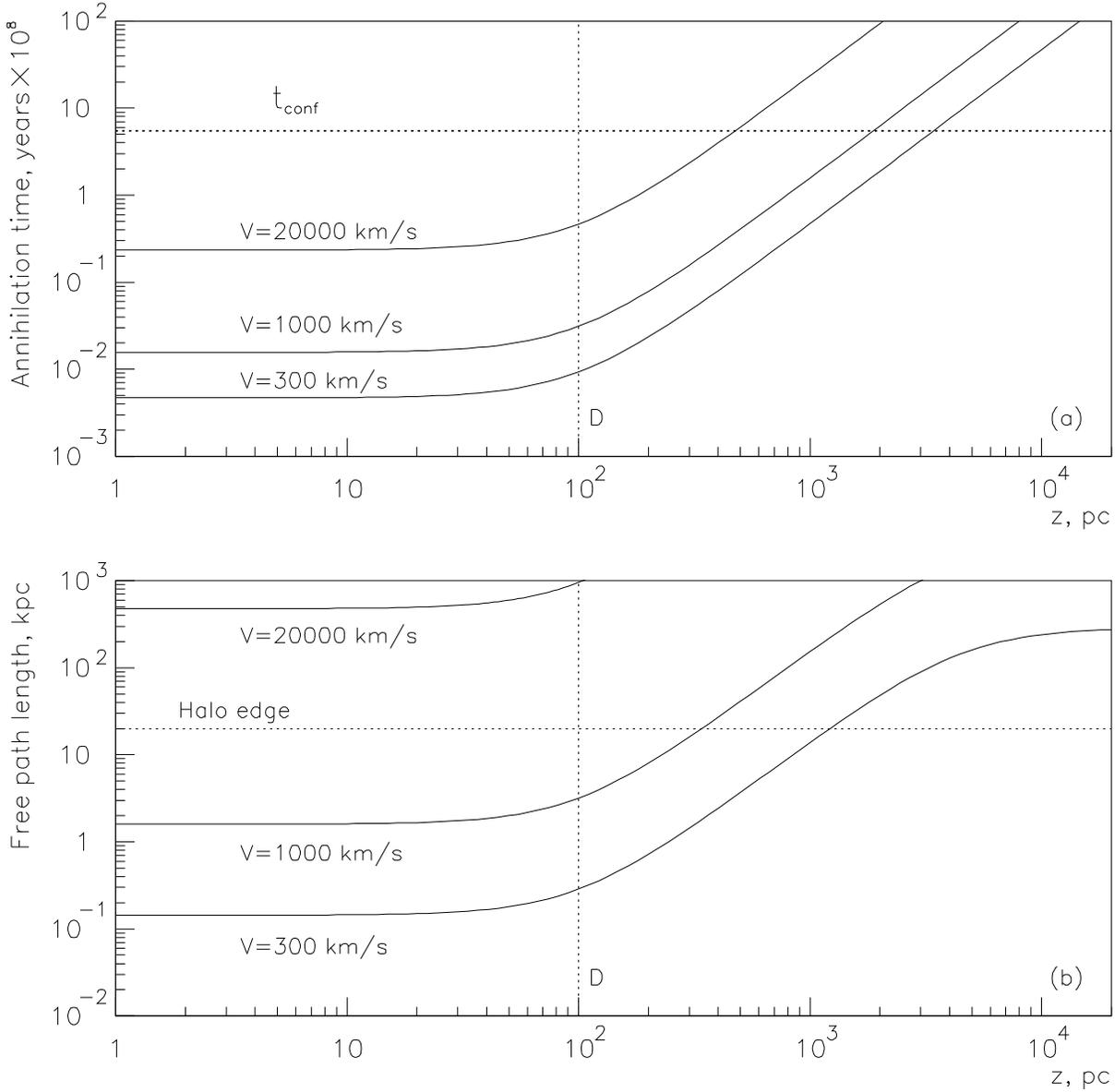,bbllx=1.5cm,bblly=5.5cm,%
bburx=19.0cm,bbury=23.0cm,clip=t,height=16.0cm}%
}}
\par
\caption{\label{aprtime}
(a) The dependence of the antiprotons annihilation time on $z$ coordinate. 
The horizontal dashed line is the antiproton confinement time in the Galaxy.
(b) The dependence of free path length of the antiprotons.
The horizontal dashed line is the halo edge $z\,=\,20$ kpc.
The curves are calculated for three values of the antiproton velocity:
$300$ km/s, $10^3$ km/s and $2\cdot 10^4$ km/s.
Vertical dashed line shows the half-width of the disk $D\,=\,100$ pc.
}
\end{figure}

Fig.\ref{aprtime} shows the antiproton life-time to the annihilation (a)
and the free path length of the antiprotons (b) versus their distance
of the galactic plane, $z$ for three values of the antiproton velocity.
In the stationary case to compensate the annihilation of the antiprotons 
with matter gas the number density of the antiprotons
must satisfy the equation:

\beq{aprsrc}
\frac{d^2 n_{\bar p}}{dE\,dt}\ =\ I_{\bar p}(E)
\ -\ v\,\sigma(v)\,n_H\,\frac{dn_{\bar p}}{dE}\,.
\eeq

The solution of this equation is:

\beq{aprsol}
\frac{dn_{\bar p}}{dE}\ =\ I_{\bar p}(E)\,t_{ann}(E)
\,\left ( 1\,-\,e^{-t/t_{ann}}\right )\,,
\eeq

\noindent with $t_{ann}\,=\,\left [v\,\sigma(v)\,n_H\right ]^{-1}$ 
being the life-time of the antiprotons relative to the annihilation.

From Fig.\ref{aprtime}(a) we can conclude that for antiprotons
with velocities 10$^3$ km/s (stellar wind) the
confinement time in the halo, starting from distancies $z\,\sim\,2$ kpc,
is less than their annihilation time. 
Thus, from \Eq{aprsol} we obtain for the halo:

\beq{tstorh}
n(E) \ \approx\ I_{\bar p}(E)\,T_{conf}\,.
\eeq

In the gaseous disk the situation is just opposite. 
The antiprotons annihilate with high rate and their
life-time to the annihilation is much less than the time necessary
to escape the Galaxy volume. 

Other words, the antiprotons are storaging in the halo during
the confinement time $\approx\,5\cdot 10^8$ yrs increasing
the gamma flux by factor $T_{conf}$.
We can also conclude that during large confinement time
the antiprotons are being spread over the halo with constant number
density not depending on the position of the antistars cluster
and under usual acceleration mechanisms in the halo their energy
spectrum comes to the stationary form.
Additionally from Fig.\ref{aprtime}(a) we see that the ''storaging'' volume 
is order of the volume of the halo $V_{halo}\,=\,4\pi R_{halo}^3/3$ 
when the region with $T_{conf}\,>>\,T_{ann}$ is restricted
by $\vert z\vert\,\le\,2$ kpc. Thus intensive annihilation takes place 
within the volume $V_{ann}\,\approx\,\pi R_{halo}^2\,4\,kpc$.
The ratio of these two volumes is order of

\bma
\frac{V_{ann}}{V_{halo}}\ \sim\ \frac{4\,kpc}{4/3\,R_{halo}}
\ \le\ 20\%
\ema

\noindent and the annihilation of the antiprotons in the gaseous disk
practically does not affect the number density
of the antiprotons in the Galaxy as whole.

The above consideration provides quasi-stationary distribution 
of antimatter in the halo and, as results, constant
number density of the antiprotons in the galactic halo.
Fig.\ref{aprtime}(b) shows $z$ dependence of free path length 
of the antiprotons at three values of their velocity.

\section{Diffuse gamma flux.}

The gamma flux arriving from the given direction is defined
by the well known expression:

\beq{jgamma}
J_{\gamma}(E_{\gamma})\ =\ \int_0^L\,dl\,\psi (E_{\gamma},r,z).
\eeq

The integration must be performed 
up to the edge of the halo $L\,=\,-\alpha_x\,R_{\odot}\,+\,
\sqrt{R_{halo}^2-R_{\odot}^2\,\left (1-\alpha_x^2\right )}$
with $\alpha_x$ being the cosine of the line-of-sight to the $x$ axis,
directed from the Galaxy Center to the Sun and lying in the plane
of the Solar orbit.

Function $\psi (E_{\gamma})$ in \Eq{jgamma} is the intensity of gamma sources 
along the observation direction $l$ in assumption of isotropic 
distribution of gamma emission. This function is defined as:

\beq{srcgamma}
\begin{array}{lll}
\psi (E_{\gamma},r,z) & = & \int_{E_{min}}^{\infty}\,dE
\,v(E)\,\sigma_{ann}(E)\,n_H(r,z)\,n_{\bar p}(E,r,z)\,W(E_{\gamma};E)\\
\\
W(E_{\gamma};E) & = &\frac{dn_{\gamma} (E_{\gamma};E)}{dE_{\gamma}\,dO}\,.
\end{array}
\eeq

To simulate the gamma energy spectrum and angular distribution 
$W(E_{\gamma};E)$ we used the Monte Carlo technics.
The experimental data \cite{Backenstoss83} on the $\bar pp$
annihilation at rest (see Table) have been used
to simulate the probabilities of different final states.
In practice, the approximation of the annihilation at rest is valid 
with very good accuracy up to laboratory momenta of the incoming 
antiprotons about $0.5$ GeV because at these laboratory momenta 
the kinetic energy of the antiproton is still order of magnitude 
less than the twice antiproton mass.
The simulation of the distributions of final state particles 
has been performed according to phase space in the center-of-mass 
of the $\bar pp$ system. PYTHIA 6.127 package \cite{Pythia6} 
has been used to perform the subsequent decays of all unstable 
particles. Momenta of stable particles 
($e^{\pm}$, $p/\bar p$, $\mu^{\pm}$, $\gamma$ and neutrinos) have been
boosted in the laboratory reference frame.
The resulting average number of $\gamma$'s per annihilation is

\bma
<n_{\gamma}>\ =\ \int\,d\Omega\,dE_{\gamma}\,W(E_{\gamma};E)\ =\ 3.93\,\pm\,0.24
\ema

\noindent and agrees with experimental data.

In the stationary case we can put that annihilation rate in the halo 
is being constantly compensated by the permanent source 
of the antiprotons.
But, owing to the fact that the antiprotons annihilation rate 
in the gaseous disk is much greater than in halo, 
we need to take into account the dependence of the antiproton density 
on $z$ coordinate. Fig.\ref{aprtime}(b) demonstrates that free path length 
of the slowest antiprotons is comparable with half-width of the disk $D$.
To take this effect into account we have to consider the annihilation 
with disk gas. For given value of $z$ we have:

\beq{dskann}
\frac{dn_{\bar p}(z,E)}{dz}\ =\ \sigma_{ann}(E)\,\Delta_H(z)
\,n_{\bar p}(z,E)\,.
\eeq

The differential equation \Eq{dskann} can be easily solved and results
the following antiproton number density distribution along $z$ axis:

\beq{aprdens}
n_{\bar p}(z,E)\ =\ n_0\,\exp
\left \{ -\sigma_{ann}(E)\, \int_z^{z_{max}}\, dz'
\,\Delta_H(z') \right \},
\eeq

\noindent where, $z_{max}\,=\,L\,\alpha_z$ is the maximal value of $z$ 
coordinate, defined by the edge of the halo,
and $n_0$ is the antiproton number density far from the disk.

The next point we need to consider is the antiproton energy spectrum.
As it will be shown further, the stellar wind from antistars has to give more
significant contribution in the antimatter pollution from the anticluster.
The original distribution of the stellar wind particles has a Gaussian form
peaking at velocities $v\,\approx\,500$ km/s \cite{Ellison98}.
The interplanetary shocks accelerate emitted particles and the resulting
stellar cosmic rays flux becomes proportional to
$J_{SW}\,\sim\,v\,E_{kin}^{-2}$ in the range
of kinetic energies up to $\sim\,100$ MeV \cite{Ellison98}.
Additional acceleration occurs in the interstellar plasma and,
as we believe, produces the observable spectrum of the galactic cosmic rays
$\sim\,v\,E_{kin}^{-2.7}$.
Both the acceleration mechanisms are being defined by the collisionless 
shocks in interplanetary or Galaxy plasmas and are charge-independent.
One has to take into account also the relative movement of the hypothetical
antistars cluster with velocity $\sim\,300$ km/s as well as the similar
velocities of the matter gas defined by the gravitational field
of the Galaxy. Thus, one can expect that minimal relative velocity
of the antiprotons from (anti)stellar wind and the matter gas is
something about $v_{min}\,\approx\,600-700$ km/s.
Following the above consideration, we chose the antiproton spectrum
in the halo (far from regions with high matter gas density)
to be similar to the galactic cosmic-rays proton spectrum in the whole range 
of the antiproton energies:

\beq{aprflux}
n_{\bar p}(E,z>>D)\ \sim\ \left (\frac{1\,\mbox{GeV}}{E_{kin}}\right )^{2.7},
\eeq

\noindent with the normalization:

\bma
\int_{E_{min}}^{\infty}\,n_{\bar p}(E,z>>D)\,dE\ =\ n_0\,.
\ema

Actually, reasonable variation of the form of the antiproton flux 
does not affect significantly the total normalization and changes only 
the gamma spectrum at higher energies. 
The main contribution in the integrated antiproton number density
comes from the slowest antiprotons owing to fast growth
of the annihilation cross section with decrease of the velocity.
We don't consider in present paper the contribution in the gamma flux
from the annihilation of the secondary antiprotons produced in the collisions
of the cosmic-ray protons with interstellar gas. This effect must give
the main contribution at higher energies of gammas and needs careful investigation
of the deceleration mechanisms in the halo.

If we assume that all the gamma background at high galactic latitudes 
is defined by the antiproton annihilation,
we have the only free parameter in our model - the minimal velocity
of the antiprotons $v_{min}$. Therefore, for given value $v_{min}$
the integrated number density of the antiprotons in the halo $n_0$
can be found from comparison with the observational data on diffuse gamma flux.   
If we choose the minimal velocity of the antiprotons 
order of the velocity of the stellar wind, $v_{SW}\,\approx\,1000$ km/s,
being equivalent to kinetic energy of the antiprotons
$E_{kin}^{SW}\,\approx\,5.2$ keV, we obtain
the necessary integral number density of the antiprotons
$n_0$ to be equal to:

\beq{aprsw}
n_0^{SW}\ \approx\ 5.0\cdot 10^{-12}\ \mbox{cm}^{-3}.
\eeq

\begin{figure}[htb]                
\par
\centerline{\hbox{%
\psfig{figure=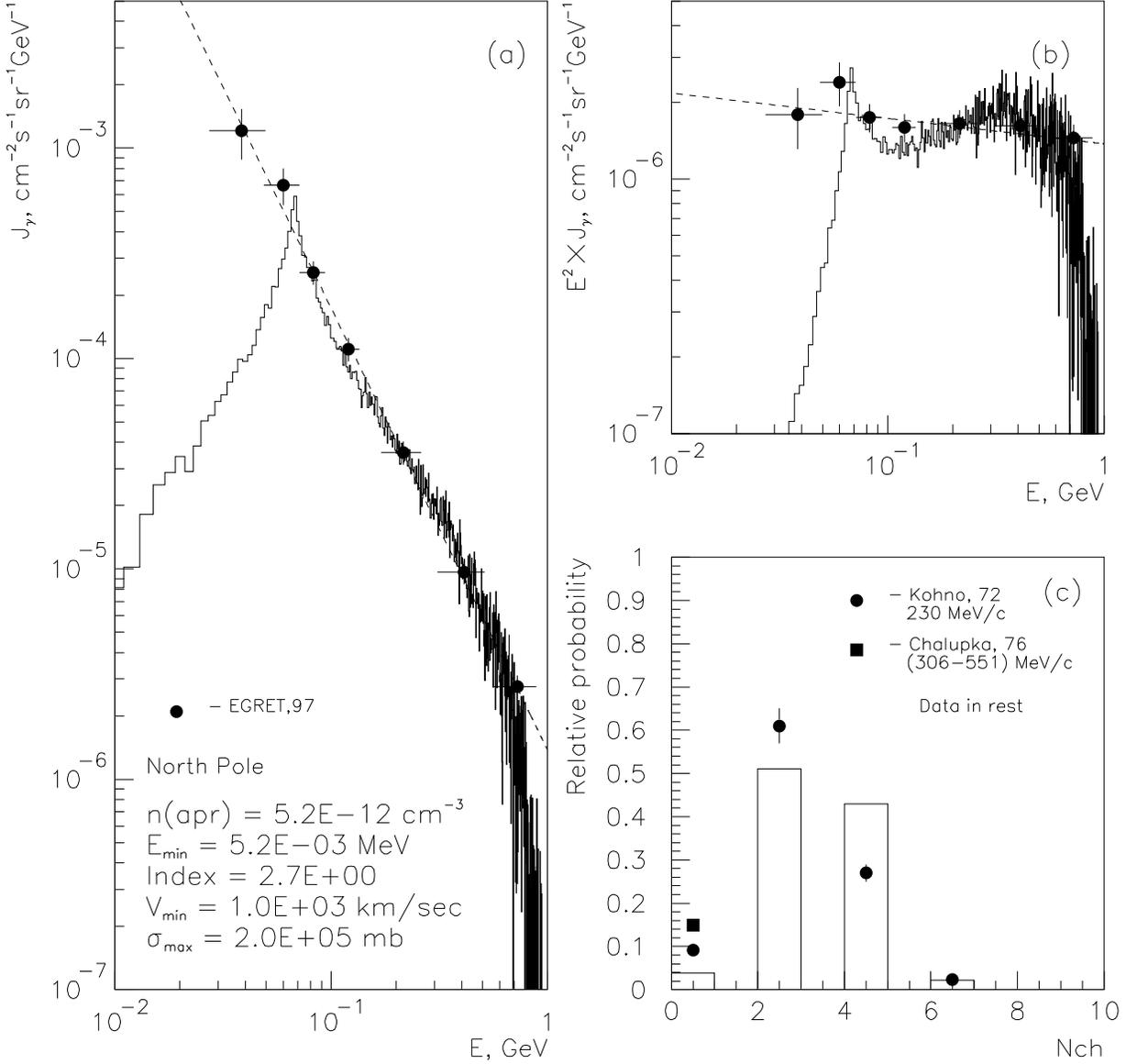,bbllx=1.5cm,bblly=5.5cm,%
bburx=19.0cm,bbury=23.0cm,clip=t,height=16.0cm}%
}}
\par
\caption{\label{gmflux}
Comparison of the calculated differential fluxes 
of $\gamma$ quanta from $\bar p/p$ annihilation for the minimal
antiproton velocity $v_{min}\,=\,10^3$ km/s
with experimental data {\em EGRET} \cite{EGRET97}
on diffuse gamma background (a,b).
The observational direction is to the North Pole of the Galaxy.
There is also shown the comparison
of the charged multiplicity distribution in the annihilation model 
described in the text with the existent experimental data (c).
Circles - \cite{Kohno72}, squares - \cite{Chaloupka76}.
}
\end{figure}

Fig.\ref{gmflux}(a,b) demonstrates the resulting differential gamma 
distribution in the Galactic North Pole direction
in comparison with EGRET data \cite{EGRET97} in the range 
$10\,\le\,E_{\gamma}\,\le\,1000$ MeV.
The peak of $\pi^0$ decay is clearly seen both in calculations as well
as in experimental distributions. Fig.\ref{gmflux}(c) shows the charged
multiplicity distribution in the annihilation model described above.
The comparison with the experimental points taken from
\cite{Kohno72,Chaloupka76} serves as additional confirmation 
of our calculations.

We also performed calculations for two other values of the minimal
velocity of the antiprotons $v_{disp}\,=\,300$ km/s
and for the velocity of the (anti)matter thrown out by the Supernovae,
$v_{SN}\,=\,2\cdot 10^4$ km/s. The respective necessary values 
of the integral antiproton number density are:

\beq{aprsn}
\begin{array}{lll}
n_0^{disp} & \approx\ & 2.0\cdot 10^{-12}\, \mbox{\nsz cm}^{-3}\\
\\
n_0^{SN} & \approx & 6.0\cdot 10^{-11}\, \mbox{\nsz cm}^{-3}\,.
\end{array}
\eeq

Thus, one can see that necessary integral antiproton density in the halo
practically linearly depends on minimal velocity of the antiprotons
in the range $300\,\le\,v\,\le\,10^4$ km/s.
Note, that the approximation about annihilation at rest is valid for all
the range of above minimal velocities and the resulting gamma spectrum
does not change its form at such a variation of $v_{min}$.


\section{Discussion and Conclusion}

Let us estimate the intensity of the antiproton source
and, as result, the total mass of the hypothetical globular
cluster of antistars for three values of the minimal antiproton velocity: 
$v_{disp}$, $v_{SW}$ and $v_{SN}$. 
The first case assumes that antiprotons have been decelerated and
travel in the halo with velocities equal to the velocity dispersion
defined by the galactic gravitational field. The second value
of $v_{min}$ is the order of the speed of the fast stellar wind and
the third case is the velocity of the particles blown off by the Supernova 
explosion without possible deceleration.
 
If we integrate over the volume of the whole halo and take into account
the antiproton storaging in the halo during the confinement time,
we obtain for the integral intensity of the antiproton source 
$\dot{M}\,\sim\,\left (n_0\,m_p\,V_{halo}\right )/t_{conf}$.
For above three variants of the minimal velocity of the antiprotons
and  $t_{conf}\,\sim\,5\cdot 10^8$ years
from \Eq{aprsw} and \Eq{aprsn} we obtain the following values 
of the necessary antiproton source intensity:

\beq{aprate}
\begin{array}{lll}
\dot{M}^{disp} & \approx & 3.0\cdot 10^{-9}\ M_{\odot}/yr\\
\\
\dot{M}^{SW} & \approx & 8.5\cdot 10^{-9}\ M_{\odot}/yr\\
\\
\dot{M}^{SN}   & \approx & 1.0\cdot 10^{-7}\ M_{\odot}/yr\\
\end{array}
\eeq

From the analogy with elliptical galaxies 
in the case of constant mass loss due to stellar wind 
one has the mass loss $10^{-12}M_{\odot }$ per Solar mass per year.
In the case of stellar wind we find for the mass of the anticluster:

\beq{mclusw}
M_{clu}^{SW}\ \approx\ 2\cdot 10^4\,M_{\odot}\,.
\eeq

To estimate the frequency of Supernova explosions in the antimatter 
globular cluster the data on such explosions in the elliptical galaxies 
were used \cite{Khlopov98}, what gives the mean time interval between 
Supernova explosions in the antimatter globular cluster $\Delta T_{SN}\,
\sim\,1.5\cdot 10^{15}\,M_{5}^{-1}$ s. For $M_{5}\,>\,1$ this interval
is smaller than the period of the orbital motion of the cluster, 
and one can use the stationary picture considered above with the change 
of the stellar wind mass loss by
the $\dot{M}\,\sim\,f_{SN}\cdot M_{SN}$, where $f_{SN}\,=\,6\cdot 10^{-16}
\,M_{5}$ s$^{-1}$ is the frequency of Supernova explosions and $M_{SN}\,=\,
1.4\,M_{\odot}$ is the antimatter mass blown off in the explosion.
Following the theory of Supernova explosions in old star  populations 
only the supernovae of the type I (SNI) take place, in which no hydrogen 
is observed in the expanding shells. In strict analogy with the matter SNI 
the chemical composition of the antimatter Supernova shells should include 
roughly half of the total ejected mass in the internal anti-iron shell 
with the velocity dispersion $v_i\,\le\,8\cdot 10^8$ cm/s 
and more rapidly expanding $v_e\,\sim\,2\cdot 10^9$ cm/s 
anti-silicon and anti-calcium external shell.
The averaged effective mass loss due to Supernova explosions gives 
the antinucleon flux $\dot{N}\,\sim\,10^{42}\,M_{5}\,s^{-1}$, 
but this flux contains initially antinuclei with the atomic number 
$A\,\approx\,30\,-\,60$, so that the initial flux of antinuclei is equal to 
$\dot{A}\,\sim\,(2\,-\,3)\cdot 10^{40}\,M_{5}\,s^{-1}$. 
Due to the factor $\sim\,Z^2 A^{2/3}$
in the cross section the annihilation life-time of such nuclei 
is smaller than the cosmic ray life-time, and in the stationary picture 
the products of their annihilation with $Z\,<\,10$ should be considered. 
With the account for the mean multiplicity $<N>\,\sim\,8$ 
of annihilation products one obtains the effective flux 
$\dot{A}_{eff}\,\sim\,(1.5\,-\,2.5)\cdot 10^{41}\,M_{5}$ s$^{-1}$,
being an order of magnitude smaller than the antiproton flux from
the stellar wind. 

If to take the antimatter stellar wind as small as the Solar wind
$(\dot{M_{\odot}}\,=\,10^{-14}\,M_{\odot}$ yr$^{-1})$ this corresponds 
to the antiproton flux by two orders of magnitude smaller 
than one chosen above in \Eq{aprate}, and the antimatter from Supernova 
should play the dominant role in the formation of galactic gamma background. 
For the Supernova case we have for the mass of the anticluster
the value

\bma
M_{clu}^{SN}\,\approx\,4.0\cdot 10^5\,M_{\odot}\,,
\ema

\noindent which value agrees with the estimation \cite{Khlopov98}.
If we assume that significant fraction of the antiprotons from stellar
wind is decelerated up to $v_{disp}$ the respective mass 
of the globular cluster of antistars can be reduced up to

\bma 
M_{clu}^{disp}\,\approx\,7\cdot 10^{3}\,M_{\odot}.
\ema

It is necessary to make small remark. Namely, in principle,
one cannot exclude that the secondary antiprotons produced in $pp$
collisions can be decelerated in the halo magnetic fields
up to velocities order of few hundreds km/s. 
In this case they will also give contribution in the diffuse
gamma flux annihilating with the matter gas and the calculations
performed in present paper are valid in this case also.

In conclusion we can say that the hypothesis on the existence of antimatter
globular cluster in the halo of our Galaxy
does not contradict to either modern particle physics models
or observational data. 
Moreover, the Galactic gamma background measured by EGRET can be explained
by antimatter annihilation mechanism in the framework of this hypothesis.
If the mass
of such a globular cluster is of order of $10^4\div 10^5\ M_{\odot}$,
we can hope that other signatures of its existence like fluxes
of antinuclei can be reachable for the experiments in the nearest 
future.
The analysis of antinuclear annihilation cascade 
is important in the realistic estimation of antinuclear cosmic ray 
composition but seems to be much less important in its contribution 
into the gamma background as compared with the effect of antimatter 
stellar wind. This means that the gamma background and the cosmic
antinuclei signatures for galactic antimatter are complementary and
the detailed test of the galactic antimatter hypothesis is possible
in the combination of gamma ray and cosmic ray studies.

\bigskip
{\it Acknowledgements}.
The authors acknowledge the COSMION Seminar participants 
for useful discussions.
The work was partially carried out in framework 
of State Scientific Technical Programme ''Astronomy. 
Fundamental Space Research'', Section ''Cosmoparticle Physics''.
One of the authors (M.Kh.) expresses his gratitude also to COSMION-ETHZ
and AMS-EPICOS collaborations for permanent support.

\vfill\eject
\newcommand{\jour}[4]{ #1, {\bf #2}, #3 (#4)}

\newpage

\newpage

\newpage

\newpage

\newpage
\vspace*{0.5cm}

\vbox{
\centerline{Table. Relative probabilities of $\bar pp$ annihilation channels.}
\bma
\begin{array}{|l|r||l|r|}
\hline
&&&\\
\mbox{\rm Channel} & \mbox{\rm Rel. prob.}, \% 
& \mbox{\rm Channel} & \mbox{\rm Rel. prob.}, \%\\
&&&\\
\hline
&&&\\
\pi^+\pi^-\pi^0   & 3.70     & 2\,\pi^+2\pi^-\eta  & 0.60\\
\rho^-\pi^+       & 1.35     & \pi^0\rho^0 & 1.40 \\
\rho^+\pi^-       & 1.35     & \eta\rho^0 & 0.22\\
\pi^+\pi^-2\pi^0 & 9.30      & 4.99\,\pi^0  & 3.20\\
\pi^+\pi^-3\pi^0 & 23.30     & \pi^+\pi^- & 0.40\\
\pi^+\pi^-4\pi^0 & 2.80      & 2\,\pi^+2\pi^- & 6.90\\
\omega\pi^+\pi^- & 3.80      & 3\,\pi^+3\pi^- & 2.10\\
\rho^0\pi^0\pi^+\pi^- & 7.30 & K\bar K \,0.95\pi^0 & 6.82\\
\rho^+\pi^-\pi^+\pi^- & 3.20 & \pi^0\eta ' & 0.30\\
\rho^-\pi^+\pi^+\pi^- & 3.20 & \pi^0\omega & 3.45\\
2\,\pi^+2\pi^-2\pi^0 & 16.60 & \pi^0\eta & 0.84\\
2\,\pi^+2\pi^-3\pi^0 & 4.20  & \pi^0\gamma & 0.015\\
3\,\pi^+3\pi^-\pi^0 & 1.30   & \pi^0\pi^0 & 0.06\\
\pi^+\pi^-\eta  & 1.20 &&\\
&&&\\
\hline
\end{array}
\ema
}

\end{document}